\documentclass[twocolumn,letterpaper]{article}

\renewenvironment{abstract}{%
      \list{}{\relax\small
      \leftmargin=0.25cm
      \labelwidth=0cm
      \listparindent=0cm
      \itemindent\listparindent
      \rightmargin\leftmargin}\item[\hskip\labelsep
                                    \bfseries\abstractname]}
{\endlist\bigskip}
\renewcommand{\abstractname}{Abstract.}

\usepackage[latin1]{inputenc}
\usepackage{booktabs}
\usepackage{varioref}
\usepackage{graphicx}
\usepackage{rotating}
\usepackage{xspace}

\newcommand{\captionfonts}{\small}

\makeatletter  %
\long\def\@makecaption#1#2{%
  \vskip\abovecaptionskip
  \sbox\@tempboxa{{\captionfonts {\bf #1.} #2}}%
  \ifdim \wd\@tempboxa >\hsize
    {\captionfonts {\bf #1.} #2\par}
  \else
    \hbox to\hsize{\hfil\box\@tempboxa\hfil}%
  \fi
  \vskip\belowcaptionskip}
\makeatother   %

\def\fps@figure{htbp}
\def\fnum@figure{\figurename\thefigure}
\renewcommand{\figurename}{Figure}

\newenvironment{narrow}{\begin{list}{}{\small \leftmargin 16pt \rightmargin 16pt}\item[]}{\end{list}}
\newenvironment{ind}{\begin{list}{}{\leftmargin 24pt \parsep 0pt \parskip 0pt \itemsep 0pt \topsep 0pt}\item[]}{\end{list}}

\newcommand{\fun}[1]{\textit{\textbf{#1}}}

\newcommand{\const}[1]{\textsc{#1}}
\newcommand{\keyw}[1]{\textbf{#1}}
\newcommand{\var}[1]{\textit{#1}}
\newcommand{\type}[1]{{#1}}

\newcommand{\rem}[1]{\textit{-- #1}}

\newcommand{\hide}[1]{}

\newcommand{\visb}[1]{\textbf{#1}}

\newcommand{\la}{$\langle$}
\newcommand{\ra}{$\rangle$}
\newcommand{\N}{$_{\rm N}$}
\newcommand{\F}{$_{\rm F}$}

\begin{document}

\title{\bf \vspace{-0cm} \Large A Single-Instance Incremental SAT Formulation of Proof- and Counterexample-Based Abstraction\\[0.3cm]}
\author{
\begin{tabular}{c@{~~~~~~~~~~~~~~~}c}
Niklas Een, Alan Mishchenko                & Nina Amla \\[8pt]
\small EECS Department                     & \small Cadence Research Labs \\[-2pt]
\small Univ.~of California, Berkeley, USA. & \small Berkeley, USA.
\end{tabular}
}

\date{}
\maketitle

\thispagestyle{empty}

\noindent
\begin{abstract}
This paper presents an efficient, combined formulation of two widely used abstraction
methods for bit-level verification: \emph{counterexample-based abstraction (CBA)} and
\emph{proof-based abstraction (PBA)}. Unlike previous work, this new method is
formulated as a single, incremental SAT-problem, interleaving CBA and PBA to
develop the abstraction in a bottom-up fashion. It is argued that the new method is
simpler conceptually and implementation-wise than previous approaches. As an
added bonus, proof-logging is not required for the PBA part, which allows for a
wider set of SAT-solvers to be used.
\end{abstract}

\vspace{-20pt}
\section{Introduction}

\noindent
Abstraction techniques have long been a crucial part of successful verification flows.
Indeed, the success of SAT-solving can largely be attributed to its inherent ability
to perform localization abstraction as part of its operations.
For this reason so called bug-hunting, or BMC, methods can often be applied on a full
design directly, thereby deferring the abstraction work to the SAT-solver. However,
computing an abstraction explicitly is often more useful for hard properties that
require a mixture of different transformation and proof-engines to complete the
verification.

In our formulation, both CBA and PBA compute a localization in the form of a set of
flops. An abstracted flop is in essence replaced by a primary input (\emph{PI}), thus
giving more behaviors to the circuit. Both methods work by analyzing, through the use
of SAT, a $k$-unrolling of the circuit. However, they differ as follows:
\begin{itemize}
\item CBA works in a bottom-up fashion, starting with an empty abstraction (all flops are
  replaced by PIs) and adding flops to refute the counterexamples as they are
  enumerated for successively larger $k$.
\item PBA, in contrast, considers the full design and a complete refutation of all
  counterexamples of depth $k$ (in the form of an UNSAT proof). Any flop not
  syntactically present in the proof of UNSAT is abstracted.
\end{itemize}
The two methods have complementary strengths: CBA by virtue of being bottom-up is very
fast, but may include more flops than necessary. PBA on the other hand does a more
thorough analysis and almost always gives a tighter abstraction than CBA, but at the cost
of longer runtime.

In this work, it is shown how the two methods can be seamlessly combined by applying PBA,
not on the full design, but on the latest abstraction produced by CBA\@.  This solution has
a very elegant incremental SAT formulation, which results in a simple, scalable algorithm
that has the strength of both methods.

In the experimental section it is shown how a design with 40,000 flops and 860,000
\const{And}-gates is localized to a handful of flops in just 4 seconds (much faster than
any previous method), and how this abstraction is instantaneously solved by the
interpolation-based proof-engine \cite{interpolation:03}, whilst the original
unabstracted design took 2 minutes to verify, despite the inherent localization ability
of interpolation.

\section{Related Work}

\noindent
Counterexample-based abstraction was first introduced by Kurshan in \cite{kurshan:94} and
further developed by Clarke et.~al.\ in \cite{cba:02}. Proof-based abstraction was coined
by McMillan \cite{pba:03}, and independently proposed by Gupta et.~al\ in
\cite{guptapba:03}.

The work most closely related to ours is Gupta's work of \cite{guptapba:03} and McMillan
et.~al's work of \cite{hybrid:04}.
In both approaches, abstract counterexamples are concretized using a SAT-solver. When
concretization fails, the UNSAT proof guides the abstraction refinement. Our work does
not rely on a SAT-solver to refute counterexamples, but instead uses a simpler and more
scalable method based on ternary simulation (section~\ref{cexrefine}).

Gupta's approach does not rely on BDD reachability to produce abstractions; although BDDs
are used to form a complete proof-procedure. Like our method, it tries to limit the
amount of logic that is put into the SAT-solver when unrolling the circuit, thereby
improving scalability.
It differs, though, in that the initial unrolling is done on the concrete design (our
method starts with an empty abstraction), and that PBA is used to shrink the abstract model
in a more conservative manner, requiring the PBA result to stabilize over several iterations.

The work of McMillan et.~al.\ mixes PBA for refuting all counterexamples of length $k$
with proof-analysis of counterexamples from the BDD engine, refuting individual (or small sets
of) counterexamples. Unlike Gupta's %
work, BDDs are an integral part of the abstraction computation.

The approach proposed in this paper differs further from previous work in that it does
not constitute a complete proof-procedure.  There are many different ways of using an
abstraction method as part of a verification flow.
A simple use-model would be: Run the abstraction computation until some resource limit is
reached, then output the best abstraction found so far and put the method on hold. If
the abstraction turns out not to be good enough for the downstream flow, resume
abstraction computation with a higher resource limit, and produce a more refined
abstraction. Obviously, this use-model can be further improved by multi-threading
on a multi-core machine.

In the experimental evaluation, we choose to pass abstractions to an
interpolation-based proof-engine. This particular setup relates to the work of
\cite{Li:06} and \cite{TACAS:07}.

\section{Assumptions and Notation}\label{prelim}

\noindent
In the presentation, the following is assumed:
\begin{itemize}
\item The design is given as a set of next-state functions expressed in terms of current state
  variables (flops) and primary inputs (PIs).
\item The design has only one property, which is a safety property.
\item All flops are initialized to zero, and are running on the same clock (hence acting as
unit delays).
\end{itemize}
It is further assumed that the logic of the next-state functions is represented as a
combined And-Inverter-graph, with the single property being the output of a particular
\const{And}-gate. As customary, the negation of the property is referred to as the
\var{bad} signal.

An ``abstraction'' is identified with a set of flops. If a flop is not part of the
abstraction, it is treated as a PI in the abstract model of the design. By this
semantics, adding a flop to the current abstraction means \emph{concretizing} it in the
abstract model: replace the PI by a flop and connect it to the appropriate
input signal.

\section{Algorithm}\label{algo}

\noindent
How does the proposed algorithm work? It starts by assuming the empty abstraction, treating
all flops as PIs. It then inserts one time-frame of the design into the SAT-solver, and
asks for a satisfying assignment that produces \const{True} at the \var{bad} signal. The
SAT-solver will come back SAT%
\footnote{The very first query only comes back ``UNSAT'' if the property holds
  combinationally, a corner case we ignore here.}
and the counterexample is used to concretize some of the flops (=~CBA). When enough flops
have been concretized, the SAT-problem becomes UNSAT, which means that all counterexamples
of length~0 have been refuted (unless there is a true counterexample of length zero). The
algorithm can now move on to depth~1, but before doing so, any flop that did not occur in
the UNSAT proof is first removed from the abstraction (=~PBA). The procedure is
repeated for increasing depths, resulting in an incremental sequence of SAT calls that looks
something like
\begin{narrow}
\emph{depth 0:}~~SAT, SAT, SAT, SAT, \visb{UNSAT} \\
\emph{depth 1:}~~SAT, \visb{UNSAT} \\
\emph{depth 2:}~~SAT, SAT, SAT, \visb{UNSAT} \\
$\vdots$
\end{narrow}
with each sequence of calls at a given depth ending in an ``UNSAT'' result that prunes
the abstraction built up by analyzing the preceding ``SAT'' counterexamples.

The algorithm terminates in one of two ways:
either (i) CBA comes back with the same set of flops as were given to it, which means we
have found a true, justified counterexample, or (ii) it runs out of resources for doing
abstraction and stops.
The resulting abstraction is then returned to the caller to be used in the next
step of the verification process.

\subsection{{Counterexample-based refinement}}\label{cexrefine}

\noindent
Assume that for the current abstraction $A$ the last call to SAT returned a
counterexample of length $k$.  The counterexample is then analyzed and refined by the
following simple procedure%
\footnote{This procedure (implemented by Alan Mishchenko in ABC \cite{AlanABC}),
  has been independently discovered by one of our industrial collaborators, and
  probably by others too.  A similar procedure is described in \cite{ketchum:01}. }
in order to refute it:
\begin{narrow}
\visb{CBA refinement.}
Loop through all flops not in $A$. Replace the current value of the counterexample
with an $X$ (the undefined value) and do a three-valued simulation. If the $X$ does not
reach the \var{bad} signal, its value is unimportant for the justification of the
counterexample, and the corresponding flop is kept as a PI. If, on the other hand, $X$
propagates all the way to \var{bad}, we undo the changes made by that particular
$X$-propagation and add the corresponding flop to $A$.
\end{narrow}
The order in which flops are inspected does matter for the end result. It seems like a good
idea to consider multiple orders and pick the one producing the smallest abstraction.
But in our experience it does not improve the overall algorithm. The extra runtime may
save a few flops temporarily, but they are typically added back in a later iteration,
or removed by PBA anyway, resulting in the same abstraction in the end.

\subsection{{Incremental SAT}}

\noindent
Incremental SAT is not a uniquely defined concept. The interpretation
used here is a solver with the following two methods:
\begin{itemize}
\item \fun{addClause}(\var{literals}): This method adds a clausal constraint, i.e.~($p_0
  \lor p_1 \lor \ldots \lor p_{n-1}$) where $p_i \in$ \var{literals}, to the SAT-solver.
   The incremental interface allows for more clauses to be added later.
\item \fun{solveSat}(\var{assumps}): This method searches for an assignment that satisfies
  the current set of clauses under the unit assumptions \var{assumps}~=~$a_0 \land a_1
  \land \ldots \land a_{n-1}$.  If there is an assignment that satisfies all the clauses
  added so far, as well as the unit literals $a_i$, that
  model is returned. If, on the other hand, the problem is UNSAT under the given
  assumptions, \emph{the subset of those assumptions used in the proof of UNSAT} is
  returned in the form of a final conflict clause.
\end{itemize}
The extension of \fun{solveSat}() to accept a set of unit literals as assumptions, and to produce
the subset of those that were part of the UNSAT proof, can easily be added to any modern
SAT-solver.%
\footnote{Two simple things should be done: (i) the decision heuristic has to be changed
  so that the first $n$ decisions are made on the assumption literals; and (ii) if a
  conflict clause is derived that contradicts the set of assumptions, that clause has to
  be further analyzed back to the decision literals rather than the first UIP. For more details,
  please review the \fun{analyzeFinal()} method of MiniSAT \cite{MiniSAT}.}
This is in contrast to adding proof-logging, which is a non-trivial endeavor. For that reason,
the proposed algorithm is stated entirely in terms of this interface and does not rely on
generating UNSAT proofs.

\subsection{{Refinement using activation literals}}

\noindent
Unlike the typical implementation of PBA, this work uses \emph{activation literals},
rather than a syntactic analysis of resolution proofs, to determine the set of flops used
for proving UNSAT\@.
For each flop $f$ that is concretized, a literal $a$ is introduced in the
SAT-instance. As the flop input $f_{in}$ at time-frame $k$ is tied to the flop output at
time-frame $k+1$, the literal is used to activate or deactivate propagation through the
flop by inserting two clauses stating:
$$a \rightarrow (f[k+1] \leftrightarrow f_{in}[k])$$
The set of activation literals is passed as assumptions to \fun{solveSat}(), and for UNSAT
results, the current abstraction can immediately be pruned of flops missing from the
final conflict clause returned by the solver.

This PBA phase is very affordable. The same SAT-problem would have to be solved in a pure
CBA based method anyway. The cost we pay is only that of propagating the assumption
literals. Because abstractions are derived in a bottom-up fashion, with the final
abstraction typically containing just a few hundred flops, the overhead is small.

\section{Implementation}\label{impl}

\noindent
This section describes the combined abstraction method in enough detail for the
reader to easily and accurately reproduce the experimental results of the final
section. The pseudo-code uses the following conventions:
\begin{itemize}
\item Symbol \visb{\&} indicates pass-by-reference.
\item The type \visb{\type{Vec}\la \type{T}\ra} is a dynamic vector whose elements are of type \type{T}.
\item The type \visb{\type{Netlist}} is an extended And-Inverter-graph. It has the
  following gate types: \const{And}, \const{PI}, \const{Flop},
  \const{Const}. Inverters are represented as complemented edges. Flops act as
  unit delays. Every netlist \var{N}, has a special gate \var{N}.\var{True} of
  type \const{Const}.
\item The type \visb{\type{Wire}} represents an edge in the netlist. Think of it as a
  pointer to a gate plus a ``sign'' bit. It serves the same function as a \emph{literal}
  w.r.t.~a variable in SAT\@. Function \fun{sign}(\var{w}) will return
  \const{True} if the edge is complemented, \const{False} otherwise. By \var{w}$_0$ and
  \var{w}$_1$ we refer to the left and right child of an \const{And}-gate. By \var{w}$_{in}$
  we refer to the input of a flop.
\item The type \visb{\type{WSet}} is a set of wires.
\item The type \visb{\type{WMap\la T\ra}} maps wires to elements of type \type{T}.
  For practical reasons, the sign bit of the wire is \emph{not} used. For
  map \var{m}, \var{m}[\var{w}] is equivalent to \var{m}[\var{$\lnot$w}].  Unmapped
  elements of \var{m} are assumed to go to a distinct element
  \const{T\_Undef} (e.g.\ \const{lit\_Undef} for literals, or \const{wire\_Undef} for
  wires).
\item The type \visb{\type{lbool}} is a three-valued boolean that is either
\emph{true}, \emph{false}, or \emph{undefined}, represented in the code by:
\const{lbool\_0}, \const{lbool\_1}, \const{lbool\_X}.
\item Every SAT-instance \var{S} (of type \visb{\type{SatSolver}}) has a special literal
  \var{S}.\var{True} which is bound to \emph{true}. Method {\var{S}.\fun{newLit}()}
  creates a new variable and returns it as a literal with positive polarity. Clauses are
  added by {\var{S}.\fun{addClause}()} and method {\var{S}.\fun{satSolve}()} commences the
  search for a satisfying assignment.
\end{itemize}
Because the pseudo-code deals with two netlists \var{N} and \var{F}, wire-types are
subscripted \type{Wire\N} and \type{Wire\F} to make clear which netlist the wire belongs
to. The same holds for \type{WSet} and \type{WMap}.

\subsection{{BMC Traces}}


\begin{figure}[t]
\begin{center}
\framebox{ \begin{minipage}{.95\linewidth} \small
\keyw{class} \type{Trace} \{ \\[-6pt] \begin{ind}
  \hspace{-24pt} \rem{Private variables:} \\[4pt]
  \begin{tabular}{@{}l@{~}l}
  \type{Netlist}\&  & \var{N}; \\
  \type{Netlist}    & \var{F}; \\
  \type{SatSolver}  & \var{S}; \\
  \type{WSet\N}     & \var{abstr}; ~~~~~~\rem{publicly read-only} \\
  ~\\
  \type{Vec\la WMap\N\la Wire\F\ra\ra}  & \var{n2f}; \\
  \type{WMap\F\la Lit\ra}               & \var{f2s}; \\
  \type{WMap\N\la Lit\ra}               & \var{act\_lits}; \\
  \end{tabular} \\[6pt]
  \phantom{} \hspace{-24pt} \rem{Private functions:} \\[2pt]
  \begin{tabular}{@{}l@{~}l@{~}l}
  \type{Lit}    & \fun{clausify}  & (\type{Wire\F} \var{f}); \\
  \type{void}   & \fun{insertFlop}& (\type{int} \var{frame}, \type{Wire\N} \var{w\_flop}, \type{Wire\F} \var{f}); \\
  \end{tabular}
  ~\\
  \phantom{} \hspace{-24pt} \rem{Constructor:} \\[2pt]
  \fun{Trace}(\type{Netlist}\& \var{N}); \\[6pt]
  \phantom{} \hspace{-24pt} \rem{Public functions:} \\[2pt]
  \begin{tabular}{@{}l@{~}l@{~}l}
  \type{Wire\F} & \fun{insert}    & (\type{int} \var{frame}, \type{Wire\N} \var{w}); \\
  \type{void}   & \fun{extendAbs} & (\type{Wire\N} \var{w\_flop}); \\
  \type{bool}   & \fun{solve}     & (\type{WSet\F} \var{f\_disj}); \\
  \type{Cex}    & \fun{getCex}    & (\type{int} \var{depth});
  \end{tabular}
\end{ind} \};
~\\[10pt]
\keyw{class} \type{Cex} \{ $\ldots$ \}; ~~~~~~\rem{stores a counter-example}
\phantom{x} \end{minipage} }
\end{center}\vspace{-8pt}
\caption{\label{fig:Trace}
\emph{Interface of the ``Trace'' class.} The class handles the BMC unrolling of the
design \var{N}. Netlist \var{F} will store the structurally hashed
unrolling of \var{N}. SAT-solver \var{S} will store a CNF representation
of the logic in \var{F}.  }
\end{figure}

\noindent
To succinctly express the SAT analysis of the unrolled design, the class
\visb{\type{Trace}} is introduced (see Figure~\ref{fig:Trace}). It allows for incrementally
extending the abstraction, as well as lengthening the unrolled trace.
Its machinery needs the following:
\begin{itemize}
\item A reference \var{N} to the input design (read-only).
\item A set of flops \var{abstr}, storing the current abstraction. Calling \fun{extendAbs}()
  will grow this set. Calling \fun{solve}() may shrink it through its built-in PBA.
\item A netlist \var{F} to store the unrolling of \var{N} under the current abstraction.
  Gates are put into \var{F} by calling \fun{insert}(\var{frame}, \var{w}). Only the
  logic reachable from gate \var{w} of time-frame \var{frame} is inserted. For
  efficiency, netlist \var{F} is kept structurally hashed.
\item A SAT-instance \var{S} to analyze the logic of \var{F}. Calling
  \fun{solve}(\var{f\_disj}) will incrementally add the necessary clauses to model the
  logic of \var{F} reachable from the set of wires \var{f\_disj}.  The user of the class
  does not have to worry about how clauses are added; hence \fun{clausify}() is a private
  method. The SAT-solving will take place under the assumption \var{f}$_0$ $\lor$
  \var{f}$_1$ $\lor \ldots \lor$ \var{f}$_{n-1}$.  The method \fun{solve}() has two
  important side-effects:
  \begin{itemize}
  \item For satisfiable runs, the satisfying assignment is stored so that \fun{getCex}()
    can later retrieve it.
  \item For unsatisfiable runs, the flops not participating in the proof are \emph{removed} from
    the current abstraction.
  \end{itemize}
\item Maps \var{n2f} and \var{f2s}. Expression ``\var{n2f}[\var{d}][\var{w}]'' gives the
  wire in \var{F} corresponding to gate \var{w} of \var{N} in frame \var{d}. Expression
  ``\var{f2s}[\var{f}]'' gives the literal in \var{S} corresponding to gate \var{f} of \var{F}.
\item Map \var{act\_lits}. Expression ``\var{act\_lits}[\var{w\_flop}] gives the activation
literal for flop \var{w\_flop}, or \const{wire\_Undef} if none has been introduced.
\end{itemize}

\subsection{{The main procedure}}\label{mainproc}


\begin{figure*}[!hp]
\begin{center}
\framebox{ \begin{minipage}{.975\linewidth} \small
\type{WSet\N} $\uplus$ \type{Cex} \fun{combinedAbstraction}(\type{\var{Netlist}} N) \{ \begin{ind}
    \begin{tabular}{@{}l@{~}l}
    \type{Trace}  & \var{T}(\var{N}); \\
    \type{Wire\N} & \var{bad} = $\lnot$\var{N}.\fun{getProperty}(); \\
    \type{WSet\F} & \var{bad\_disj} = $\emptyset$;
    \end{tabular}\\[6pt]
    \keyw{for} (\type{int} \var{depth} = 0;;) \{ \begin{ind}
        \keyw{if} (\la\emph{reached resource limit}\ra) \begin{ind}
            \keyw{return} \var{T}.\var{abstr};
        \end{ind}
        ~\\[-6pt]
        \var{bad\_disj} = \var{bad\_disj} $\cup$ \{\var{T}.\fun{insert}(\var{depth}, \var{bad})\}; \\
        ~\\[-6pt]
        \keyw{if} (\var{T}.\fun{solve}(\var{bad\_disj})) \{  ~~~~\rem{Found counter-example; refine abstraction:} \begin{ind}
            \type{int} \var{n\_flops} = \var{T}.\var{abstr}.\fun{size}(); \\
            \fun{refineAbstraction}(\var{T}, \var{depth}, \var{bad}); \\
            \keyw{if} (\var{T}.\var{abstr}.\fun{size}() == \var{n\_flops}) ~~~~\rem{Abstraction stable $\Rightarrow$ counter-example is valid:} \begin{ind}
                \keyw{return} \var{T}.\fun{getCex}(\var{depth});
            \end{ind}
        \end{ind}\}\keyw{else} \begin{ind}
            \var{depth}++;
        \end{ind}
    \end{ind} \}
\end{ind} \}
~\\[10pt]
\type{void} \fun{refineAbstraction}(\type{Trace}\& \var{T}, \type{int} \var{depth}, \type{Wire\N} \var{bad}) \{ \begin{ind}
    \begin{tabular}{@{}l@{~}l}
    \type{Cex}                          & \var{cex} = \var{T}.\fun{getCex}(\var{depth}); \\
    \type{Vec\la WMap\N\la lbool\ra\ra} & \var{sim} = \fun{simulateCex}(\var{T}.\var{N}, \var{T}.\var{abstr}, \var{cex});
    ~~~~~~\rem{'sim[d][w]' = value if gate 'w' at frame 'd'}
    \end{tabular} \\
    \phantom{x}\\[-2pt]
    \type{WSet\N} \var{to\_add}; \\
    \keyw{for} all flops \var{w} not in \var{T}.\var{abstr} \{ \begin{ind}
        \keyw{for} (\type{int} \var{frame} = 0; \var{frame} $\le$ \var{depth}; \var{frame}++) \{ \begin{ind}
            \fun{simPropagate}(\var{sim}, \var{T}.\var{abstr}, \var{frame}, \var{w}, \const{lbool\_X}); \\[4pt]
            \keyw{if} (\var{sim}[\var{depth}][\var{bad}] == \const{lbool\_X}) \{ \begin{ind}
                \rem{'X' propagated all the way to the output; undo simulation and add flop to abstraction:} \\
                \keyw{for} (; \var{frame} $\ge$ 0; \var{frame}$-$$-$) \begin{ind}
                    \fun{simPropagate}(\var{sim}, \var{T}.\var{abstr}, \var{frame}, \var{w}, \var{cex}.\var{flops}[\var{frame}][\var{w}]);
                \end{ind}
                \var{to\_add} = \var{to\_add} $\cup$ \{\var{w}\}; \\
                \keyw{break};
            \end{ind} \}
        \end{ind} \}
    \end{ind} \} \\[4pt]
    \keyw{for} \var{w} $\in$ \var{to\_add} \begin{ind}
        \var{T}.\fun{extendAbs}(\var{w});
    \end{ind}
\end{ind} \}
~\\[10pt]
\type{Vec\la WMap\N\la lbool\ra\ra} \fun{simulateCex}(\type{Netlist} \var{N}, \type{WSet\N} \var{abstr}, \type{Cex} \var{cex}) \{ \begin{ind}
    \keyw{return} \la\emph{ternary simulate counter-example 'cex' on 'N' under abstraction 'abstr'}\ra
\end{ind} \}
~\\[10pt]
\type{void} \fun{simPropagate}(\type{Vec$\langle$ WMap\N\la lbool\ra \ra}\& \var{sim}, \type{WSet\N} \var{abstr}, \type{int} \var{frame}, \type{Wire\N} \var{w}, \type{lbool} \var{value}) \{ \begin{ind}
    \la\emph{incrementally propagate effect of changing gate 'w' at time-frame 'frame' to 'value'}\ra
\end{ind} \}
\end{minipage} }
\caption{\label{fig:Main}
\emph{Main procedure.} Function \fun{combinedAbstraction}() takes a netlist and returns
either (i) a counter-example (if the property fails) or (ii) the best abstraction
produced at the point where resources were exhausted. We leave it unspecified what
precise limits to use, but examples include a bound on the depth of the unrolling, the
CPU time, or the number of propagations performed by the SAT solver.
Function \fun{refineAbstraction}() will use the latest counterexample stored in \var{T}
(by \fun{solve}(), if the last call was SAT) to grow the abstraction. Ternary (or
$X$-valued) simulation is used to shrink the support of the counterexample. Abstract
flops that could be removed from the support (i.e.\ putting in an $X$ did not invalidate
the counterexample) are kept abstract; all other flops are concretized.
When simulating under an abstraction, abstract flops don't use the value of their input
signal, but instead the value of the counterexample produced by the SAT solver (where the
flop is a free variable).
}
\end{center} \end{figure*}

\begin{figure*}[!hp]
\begin{center}
\framebox{
\begin{minipage}{.95\linewidth} \small
\type{Trace}::\fun{Trace}(\type{Netlist}\& \var{N0}) \{ \begin{ind}
    \var{N} = \var{N0}; \\
    \var{f2s}[\var{F}.\var{True}] = \var{S}.\var{True}; 
\end{ind} \}
~\\[10pt]
\type{Wire\F} \type{Trace}::\fun{insert}(\type{int} \var{frame}, \type{Wire\N} \var{w}) \{ \begin{ind}
    \type{Wire\F} \var{ret} = \var{n2f}[\var{frame}][\var{w}]; \\
    \keyw{if} (\var{ret} == \const{wire\_Undef}) \{ \begin{ind}
        \begin{tabular}{@{}l@{~}l@{~}l}
        \keyw{if}      & (\var{w} == \var{N}.\var{True})       & \{ \var{ret} = \var{F}.\var{True}; \} \\
        \keyw{else if} & (\fun{type}(\var{w}) == \const{PI})   & \{ \var{ret} = \var{F}.\fun{add\_PI}(); \} \\
        \keyw{else if} & (\fun{type}(\var{w}) == \const{And})  & \{ \var{ret} = \var{F}.\fun{add\_And}(\fun{insert}(\var{frame}, \var{w}$_0$), \fun{insert}(\var{frame}, \var{w}$_1$)); \} \\
        \keyw{else if} & (\fun{type}(\var{w}) == \const{Flop}) & \{ \var{ret} = \var{F}.\fun{add\_PI}(); \keyw{if} (\var{w} $\in$ \var{abstr}) \fun{insertFlop}(\var{frame}, \var{w}, \var{ret}); \}
        \end{tabular} \\
        \var{n2f}[\var{frame}][\var{w}] = \var{ret};
    \end{ind}
    \} \\
    \keyw{return} \var{ret}~\^~\fun{sign}(\var{w});  ~~~~~~~~~~~~~~~~~~~~~~~~~~\rem{interpretation: (w \^~b) $\equiv$ (b ? $\lnot$w : w)}
\end{ind} \}
~\\[10pt]
\type{void} \type{Trace}::\fun{insertFlop}(\type{int} \var{frame}, \type{Wire\N} \var{w\_flop}, \type{Wire\F} \var{f}) \{ \begin{ind}
    \type{Wire\F} \var{f\_in} = (\var{frame} == 0) ? $\lnot$\var{F}.\var{True} : \fun{insert}(\var{frame}$-$1, \var{w}$_{in}$); \\
    \type{Lit} \var{p} = \fun{clausify}(\var{f\_in}); \\
    \type{Lit} \var{q} = \fun{clausify}(\var{f}); \\
    \type{Lit} \var{a} = \var{act\_lits}[\var{w\_flop}]; \\
    \keyw{if} (\var{a} == \const{lit\_Undef}) \{ \begin{ind}
        \var{a} = \var{S}.\fun{newLit}(); \\
        \var{act\_lits}[\var{w\_flop}] = \var{a}; \}
    \end{ind}
    \var{S}.\fun{addClause}(\{{$\lnot$\var{a}, $\lnot$\var{p},  \var{q}}\}); \\
    \var{S}.\fun{addClause}(\{{$\lnot$\var{a},  \var{p}, $\lnot$\var{q}}\}); ~~~~~~~~~~~~~~~~~~~~\rem{we've now added: a $\rightarrow$ (p $\leftrightarrow$ q)}
\end{ind} \}
~\\[10pt]
\type{void} \type{Trace}::\fun{extendAbs}(\type{Wire\N} \var{w\_flop}) \{ \begin{ind}
    \var{abstr} = \var{abstr} $\cup$ \{\var{w\_flop}\}; \\
    \keyw{for} (\type{int} \var{frame} = 0; \var{frame} $<$ \var{n2f}.\fun{size}(); \var{frame}++) \{ \begin{ind}
        \type{Wire\F} \var{f} = \var{n2f}[\var{frame}][\var{w\_flop}]; \\
        \keyw{if} (\var{f} != \const{wire\_Undef}) ~~~~~~~~~~~~~~~~~~~\rem{f is either undefined or a PI} \begin{ind}
            \fun{insertFlop}(\var{frame}, \var{w\_flop}, \var{f});
        \end{ind}
    \end{ind} \}
\end{ind} \}
\end{minipage} }
\caption{\label{fig:Unroll}
\emph{Unrolling the netlist.} Method \fun{insert}() will recursively add the logic
feeding \var{w} to netlist \var{F}\@.  Flops that are concrete will be traversed across
time-frames, but not abstract flops. Each flop that is introduced to \var{F} is given an
\emph{activation literal}. If this literal is set to \const{True}, the flop will connect
to its input; if it is set to \const{False}, the flop acts as a PI. Activation literals
are used to implement the proof-based abstraction, and to disable flops when the
abstraction shrinks. At frame 0, flops are assumed to be initialized to zero.
The purpose of \fun{extendAbs}() is to grow the abstraction by one flop,
adding the missing logic for all time frames.
}
\end{center} \end{figure*}

\begin{figure*}[!hp]
\begin{center}
\framebox{
\begin{minipage}{.9\linewidth} \small
\type{Lit} \type{Trace}::\fun{clausify}(\type{Wire\F} \var{f}) \{ \begin{ind}
    \type{Lit} \var{ret} = \var{f2s}[\var{f}]; ~~~~~~~~~~~~~\rem{map ignores the sign of 'f'}\\
    \keyw{if} (\var{ret} == \const{lit\_Undef}) \{ \begin{ind}
        \keyw{if} (\fun{type}(\var{f}) == \const{PI}) \begin{ind}
            \var{ret} = \var{S}.\fun{newLit}();
        \end{ind}
        \keyw{else if} (\fun{type}(\var{f}) == \const{And}) \{ \begin{ind}
            \rem{Standard Tseitin clausification} \\
            \type{Lit} \var{x} = \fun{clausify}(\var{f}$_0$); \\
            \type{Lit} \var{y} = \fun{clausify}(\var{f}$_1$); \\
            \var{ret} = \var{S}.\fun{newLit}(); \\
            \var{S}.\fun{addClause}(\{\var{x}, $\lnot$\var{ret}\}); \\
            \var{S}.\fun{addClause}(\{\var{y}, $\lnot$\var{ret}\}); \\
            \var{S}.\fun{addClause}(\{$\lnot$\var{x}, $\lnot$\var{y}, \var{ret}\});
        \end{ind} \} \\
        \var{f2s}[\var{f}] = \var{ret};
    \end{ind} \} \\
    \keyw{return} \var{ret}~\^~\fun{sign}(\var{f});
\end{ind} \}
~\\[10pt]
%
%
\type{bool} \type{Trace}::\fun{solve}(\type{WSet\F} \var{f\_disj}) \{ \begin{ind}
    \type{Lit} \var{q} = \var{S}.\fun{newLit}(); \\
    \var{S}.\fun{addClause}(\{$\lnot$\var{q}\} $\cup$ \{\fun{clausify}(\var{f}) $|$ \var{f} $\in$ \var{f\_disj}\}); \\
    ~\\[-6pt]
     \var{assumps} = \{\var{q}\} $\cup$ \{\var{act\_lits}[\var{w}] $|$ \var{act\_lits}[\var{w}] != \const{lit\_Undef} \&\& \var{w} $\in$ \var{abstr}\}; \\
    ~\\[-6pt]
    \type{bool} \var{result} = \var{S}.\fun{solve}(\var{assumps}); \\
    \begin{tabular}{@{}l@{~}l}
    \keyw{if} (\var{result}) & \la\emph{store SAT model}\ra \\
    \keyw{else} & \var{abstr} = \var{abstr} $\backslash$ \{\var{w}\ $|$ \fun{type}(\var{w}) == \const{Flop} \&\& \var{w} $\notin$ \var{S}.\var{conflict}\};
        ~~~~~~\rem{this line does PBA}
    \end{tabular} \\
    \phantom{a}\\[-2pt]
    \var{S}.\fun{addClause}(\{$\lnot$\var{q}\});    ~~~~\rem{forever disable temporary clause} \\
    \keyw{return} \var{result};
\end{ind} \}
~\\[10pt]
\type{Cex} \type{Trace}::\fun{getCex}(\type{int} \var{depth}) \{ \begin{ind}
    \keyw{return} \la\emph{use maps 'n2f' and 'f2s' to translate the last SAT model \\
    \phantom{\keyw{return} } into 0/1/X values for the PIs and Flops of frames 0..depth}\ra
\end{ind} \}
\end{minipage}
}
\caption{\label{fig:Sat}
\emph{SAT-Solving.}  Method \fun{clausify}() translates the logic of \var{F} into CNF for
the SAT-solver using the Tseitin transformation.  The above procedure
can be improved, e.g., by the techniques of \cite{selfSAT07,manoliSAT07}.
Method \fun{solve}() takes a disjunction of wires in \var{F} and searches for a
satisfying assignment to that disjunction. Because only unit assumptions can be passed to
\fun{solveSat}(), a literal \var{q} is introduced to represent the disjunction, and a
temporary clause is added. Disabling the clause afterwards will in effect remove it. The
activation literals of the current abstraction are passed together with \var{q} as
assumptions to \fun{solveSat}(). The SAT-solver will give back either a satisfying
assignment (stored for later use by \fun{getCex}()), or a conflict clause expressing
which of the assumptions were used for proving UNSAT. This set is used to perform PBA.
In computing \var{assumps}, we note that ``\&\& \var{w} $\in$ \var{abstr}'' is necessary
if PBA has shrunken the abstraction.
In the experimental section, a variant (column \emph{New'} in Table~\ref{tab:Results}) is
evaluated where PBA is not applied inside \fun{solve}(). The set of redundant flops is
still computed as above, and remembered. When the resource limit is reached, those flops that
were redundant in the final UNSAT call are removed. In essence, the variant corresponds to
an incremental CBA implementation with a final trimming of the absraction by PBA.
}
\end{center} \end{figure*}

\noindent
The main loop of the abstraction procedure is given in Figure~\ref{fig:Main}.  Trace
instance \var{T} is created with an empty abstraction. For increasing depths, the
following is done:
\begin{itemize}
\item If the SAT-solver produces a counterexample, it is analyzed (by
  \fun{refineAbstraction}()) and flops are added to the abstraction to rule out this
  particular counterexample.
\item If UNSAT is returned, the depth is increased. The \fun{solve}() method
  will have performed proof-based abstraction internally and may have
  removed some flops from the abstraction.
\end{itemize}
For each new depth explored, a new \emph{bad} signal is added to \var{bad\_disj}. This
disjunction is passed as an assumption to the \emph{solve} method of \var{T}, which means
we are looking for a counterexample where the property fail in at least one time
frame. It is not enough to just check the last time frame because of PBA.

\subsection{{Unrolling and SAT solving}}\label{unrollproc}

\noindent
Figure~\ref{fig:Unroll} details how \fun{insert}() produces an unrolling of \var{N} inside
\var{F}, and Figure~\ref{fig:Sat} describes how \fun{solve}() translates the logic of \var{F} into
clauses and calls SAT.
Great care is taken to describe accurately what is implemented,
as the precise incremental SAT formulation is important for the performance and
quality. %
For the casual reader who may not want to delve into details, the following paragraph
summarizes some properties of the implementation:

As the procedure works its way up to greater and greater depth, only the logic reachable
from the \var{bad} signal is introduced into the SAT-solver, and only flops that have
been concretized bring in logic from the preceding time-frames. Constant propagation and
structural hashing is performed on the design, although constants are not propagated
across time-frames due to proof-based abstraction (PBA). Concrete flops are guarded by
activation literals, which are used to implement PBA. One literal guards all occurrences
of one flop in the unrolling. Flops that are removed by PBA will not be unrolled in
future time-frames. However, fanin-logic from removed flops will remain in F and in the
SAT-solver, but is disabled using the same activation literals.

\section{Evaluation and Conclusions}\label{eval}

\begin{table*}[t]
\begin{center}
\newcommand{\nm}[1]{\textit{#1}}
{
\footnotesize
\begin{tabular}{@{~}l@{~}|@{~~~}r@{~~~~}r@{~~~}|@{}r@{~~~}r@{~~~}r@{~~~}r@{~~}|@{}r@{~~~}r@{~~~}r@{~~~}r@{~~}|@{}r@{~~~}r@{~~~}r@{~~~}r@{~~~}r@{~}}
\toprule
~ & ~ & ~ & \multicolumn{4}{c|}{\visb{Abstr.~Size}~~(flops)} & \multicolumn{4}{c|}{\visb{Abstr.~Time} (sec)} & \multicolumn{5}{c}{\visb{Proof~Time} (sec)}  \\[4pt]
Bench. & \#Ands & \#Flops & ~~~New & New' &  ABC &  Hyb.  & ~~~New & New' & ABC &  Hyb.  & ~~~New & New' & ABC & Hyb. & No\,Abs. \\
\midrule
\nm{T0} &  57,560 &   1,549 &     {      2} & {      4} & {      2} & {      6} &     {      0.1} & {      0.1} & {      0.3} & {      0.5} &     {      0.1} & {      0.1} & {      0.1} & {      0.2} &     {  0.4} \\
\nm{T1} &  57,570 &   1,548 &     {     15} & {     15} & {     15} & {     15} &     {      1.1} & {\bf   0.7} & {      2.3} & {      9.7} &     {      0.9} & {      0.9} & {      0.9} & {      2.8} &     {  3.5} \\
\nm{S0} &   2,351 &   1,376 &     {\bf 112} & {    157} & {    174} &        -- &     {\bf   0.1} & {      0.3} & {      2.0} &          -- &     {      8.7} & {    129.7} & {\bf   5.3} &          -- &     { 21.2} \\
\nm{S1} &   2,371 &   1,379 &     {\bf 136} & {    170} & {    167} &        -- &     {      0.1} & {      0.1} & {      0.6} &          -- &     {\bf  57.8} & {    162.9} & {    104.9} &          -- &     {188.1} \\
\nm{S2} &   3,740 &   1,526 &     {\bf  83} & {    123} & {    113} & {    187} &     {      0.3} & {\bf   0.1} & {      0.6} & {     26.0} &     {\bf   1.1} & {     37.1} & {     11.8} & {    106.7} &     {  4.3} \\
\nm{D0} &   8,061 &   1,026 &     {    107} & {    112} & {\bf 106} &        -- &     {\bf   3.0} & {      3.3} & {     15.9} &          -- &     {      6.9} & {     19.6} & {\bf   4.9} &          -- &     {  7.9} \\
\nm{D1} &   7,262 &   1,020 &     {    139} & {    139} & {    139} & {    139} &     {      1.2} & {      1.2} & {      4.4} & {\bf   0.9} &     {      0.3} & {      0.3} & {      0.3} & {      2.9} &     {  0.6} \\
\nm{M0} &  17,135 &   1,367 &     {    179} & {    179} & {    180} & {\bf 178} &     {      6.8} & {\bf   6.3} & {     18.5} & {    206.5} &     {      0.2} & {      0.2} & {      0.2} & {      6.3} &     {  0.7} \\
\nm{I0} &   1,241 &   1,104 &     {     59} & {     57} & {\bf  50} &        -- &     {      0.5} & {\bf   0.1} & {      0.6} &          -- &     {      2.0} & {      1.9} & {\bf   0.7} &          -- &     {  5.8} \\
\nm{I1} & 395,150 &  25,480 &     {     24} & {     21} & {     21} & {     33} &     {      5.5} & {      1.3} & {\bf   1.1} & {     16.3} &     {      0.0} & {      0.0} & {      0.0} & {      0.3} &     { 22.1} \\
\nm{I2} &   5,589 &   1,259 &     {     45} & {\bf  44} & {     51} &        -- &     {      1.5} & {\bf   0.5} & {      1.5} &          -- &     {      6.2} & {\bf   5.7} & {      6.8} &          -- &     { 18.0} \\
\nm{I3} &   5,616 &   1,259 &     {     49} & {\bf  47} & {     52} &        -- &     {      1.2} & {\bf   0.4} & {      1.5} &          -- &     {\bf   5.9} & {      6.5} & {      6.2} &          -- &     { 19.1} \\
\nm{I4} & 394,907 &  25,451 &     {     79} & {\bf  72} & {    100} &        -- &     {     64.3} & {\bf  19.3} & {     30.9} &          -- &     {\bf   5.1} & {     15.0} & {     17.9} &          -- &     {   --} \\
\nm{I5} &   5,131 &   1,227 &     {     49} & {     44} & {\bf  38} & {     59} &     {      0.5} & {\bf   0.1} & {      0.4} & {    202.2} &     {      2.2} & {\bf   0.2} & {      0.4} & {     20.2} &     {  1.6} \\
\nm{A0} &  35,248 &   2,704 &     {\bf  61} & {     68} & {     95} & {     81} &     {      1.8} & {\bf   1.6} & {      6.3} & {      6.9} &     {     18.9} & {\bf  12.0} & {     35.7} & {     18.3} &     { 43.2} \\
\nm{A1} &  35,391 &   2,738 &     {     56} & {     56} & {     62} & {     83} &     {      2.3} & {\bf   1.7} & {      4.9} & {     11.6} &     {     15.7} & {     13.1} & {     31.1} & {\bf   6.9} &     { 29.5} \\
\nm{A2} &  35,261 &   2,707 &     {      8} & {      8} & {     18} & {     24} &     {      0.1} & {      0.1} & {      0.2} & {      0.8} &     {      0.0} & {      0.0} & {      0.0} & {      0.2} &     {  0.6} \\
\nm{A3} &  35,416 &   2,741 &     {\bf  59} & {     70} & {     79} & {     83} &     {\bf   2.2} & {      2.4} & {      7.9} & {    104.0} &     {     21.2} & {\bf  11.5} & {     79.0} & {     12.3} &     { 52.2} \\
\nm{A4} &  35,400 &   2,741 &     {\bf  63} & {     65} & {     67} & {    101} &     {      2.5} & {\bf   2.1} & {      4.4} & {     34.4} &     {\bf  11.9} & {     20.0} & {     36.2} & {     12.1} &     { 34.6} \\
\nm{F0} & 863,248 &  40,849 &     {      3} & {      3} & {      3} &        -- &     {\bf   1.0} & {      2.0} & {      3.5} &          -- &     {      0.0} & {      0.0} & {      0.0} &          -- &     { 48.2} \\
\nm{F1} & 863,251 &  40,850 &     {      4} & {      8} & {      4} &        -- &     {\bf   1.5} & {      4.7} & {      7.0} &          -- &     {      0.0} & {      2.2} & {      0.0} &          -- &     {100.6} \\
\nm{F2} & 863,254 &  40,851 &     {      5} & {      9} & {      5} &        -- &     {\bf   3.9} & {      6.1} & {      9.4} &          -- &     {      0.0} & {      2.4} & {      0.0} &          -- &     {110.1} \\
\bottomrule
\end{tabular} \vspace{-0.5cm}
}
\end{center}
\caption{\label{tab:Results}
\emph{Evaluation of abstraction techniques.}
Four implementations of hybrid counterexample- and proof-based abstraction were applied
to 22 benchmarks of more than 1000 flops, all for which the property holds. In
\emph{New'}, PBA was only applied to the final iteration (to be closer to the ABC
implementation).  The first section of the table shows the size of the designs. The
second section shows, for each implementation, the size of the smallest abstraction it
produced that was good enough to prove the property. The third and fourth sections show
the time to compute the abstraction, and the time to prove the property using
interpolation based modelchecking, with the very last column showing interpolation on the
original unabstracted design. Benchmarks with the same first letter denote different
properties of the same design. Timeout was set to 500 seconds.
}
\end{table*}

\noindent
The method of this paper was evaluated along two dimensions: (i) how does the
new abstraction procedure fare in the simplest possible verification flow, where a complete
proof-engine (in this case interpolation \cite{interpolation:03}) is applied to its
result versus applying the same proof-engine without any abstraction; and (ii) how does it
compare to previous hybrid abstraction methods---in our experiments, the implementation of
CBA and PBA inside ABC \cite{AlanABC}, and the hybrid method of McMillan
et.~al.\ \cite{hybrid:04}.

The examples used were drawn from a large set of commercial benchmarks by focusing on
designs with local properties containing more than 1000 flops.\footnote{
In other words, we've
picked examples for which abstraction should work well.  There are many verification problems
where abstraction is \emph{not} a useful technique, but here we investigate
cases where it is. }
Experiments were run on
an 2~GHz AMD Opteron, with a timeout of 500 seconds.  The results are presented in
Table~\ref{tab:Results}.

For all methods, the depth was increased until an abstraction good enough to prove the
property was found. ABC has a similar CBA implementation to the one presented in this
work (based on ternary simulation), but restarts the SAT-solver after each
refinement. ABC's PBA procedure is separate from CBA, so we opted for applying it once at
the end to trim the model returned by CBA.  This flow was also simulated in our new
algorithm by delaying the PBA filtering until the final iteration (reported in column
\emph{New'}).  This approach is often faster due to the fewer CBA refinement steps
required, but there seems to be a quality/effort trade-off between applying PBA at every
step, or only once at the end. In particular for the \emph{S} series, interleaved CBA/PBA
resulted in significantly smaller abstractions.  We have observed this behavior on other
benchmarks as well.

The McMillan hybrid technique was improved by replacing BDDs with interpolation, which
led to a significant and consistent speedup. However, our new method, and the similar
techniques of ABC, still appear to be superior in terms of scalability.  This is most likely
explained by the expensive concretization phase of the older method, which requires the
full design to be unrolled for the length of the counterexample.

The effect of an incremental implementation can be seen by comparing columns \emph{New'}
and \emph{ABC}. We have observed that the speedup tends to be more significant for
harder problems with higher timeouts.

The overall conclusion is that small abstractions help the proof-engine. However, there
are cases where a tighter abstraction led to significantly longer runtimes than a looser
one (although that effect did not manifested itself in this benchmark set). This can
partly be explained by the underlying random nature of interpolant-based model checking,
but it should also be recognized that replacing flops with PIs introduces more
behaviors, which means the SAT-solver has to prove a more general theorem. Occasionally
this can be detrimental, and offset the benefit of the reduced amount of logic that needs
to be analyzed. Altogether, it emphasizes that abstraction should be used in good
orchestration with other verification techniques.

\section{Acknowledgments}

\noindent
This work was supported in part by SRC contracts 1875.001 and 2057.001, NSF contract
CCF-0702668, and industrial sponsors: Actel, Altera, Calypto, IBM, Intel, Intrinsity,
Magma, Mentor Graphics, Synopsys (Synplicity), Tabula, Verific, and Xilinx.

\clearpage

\bibliographystyle{plain}
\bibliography{paper}

\begin{thebibliography}{10}

\bibitem{hybrid:04}
N.~Amla and K.~McMillan.
\newblock {\bf A Hybrid of Counterexample-based and Proof-based Abstraction}.
\newblock In {\em FMCAD}, 2004.

\bibitem{TACAS:07}
N.~Amla and K.~McMillan.
\newblock {\bf Combining Abstraction Refinement and SAT-based Model Checking}.
\newblock In {\em TACAS}, 2007.

\bibitem{cba:02}
P.~Chauhan, E.~Clarke, J.~Kukula, S.~Sapra, H.~Veith, and D.~Wang.
\newblock {\bf Automated Abstraction Refinement for Model Checking Large State
  Spaces Using SAT-based Conflict Analysis}.
\newblock In {\em FMCAD}, 2002.

\bibitem{selfSAT07}
N.~Een, A.~Mishchenko, and N.~Sorensson.
\newblock {\bf Applying Logic Synthesis for Speeding Up SAT}.
\newblock In {\em SAT07}, volume 4501 of {\em LNCS}, 2007.

\bibitem{MiniSAT}
Niklas Een and Niklas S\"orensson.
\newblock {\bf The MiniSat Page}.
\newblock {\em http://minisat.se}.

\bibitem{AlanABC}
Berkeley Logic~Synthesis Group.
\newblock {\bf ABC: A System for Sequential Synthesis and Verification}.
\newblock {\em http://www.eecs.berkeley.edu/\~{}alanmi/abc/}, v00127p.

\bibitem{guptapba:03}
A.~Gupta, M.~Ganai, Z.~Yang, and P.~Ashar.
\newblock {\bf Iterative Abstraction Using {SAT}-based {BMC} with Proof
  Analysis}.
\newblock In {\em ICCAD}, 2003.

\bibitem{kurshan:94}
R.~P. Kurshan.
\newblock {\bf Computer-Aided-Verification of Coordinating Processes}.
\newblock In {\em Princeton Univ.\ Press}, 1994.

\bibitem{Li:06}
B.~Li and F.~Somenzi.
\newblock {\bf Efficient Abstraction Refinement in Interpolation-Based
  Unbounded Model Checking}.
\newblock In {\em TACAS}, 2006.

\bibitem{interpolation:03}
K.~McMillan.
\newblock {\bf Interpolation and SAT-based Model Checking}.
\newblock In {\em CAV}, 2003.

\bibitem{pba:03}
K.~McMillan and N.~Amla.
\newblock {\bf Automatic Abstraction without Counterexamples}.
\newblock In {\em TACAS}, 2003.

\bibitem{manoliSAT07}
D.~Vroon P.~Manolios.
\newblock {\bf Efficient Circuit to CNF Conversion}.
\newblock In {\em SAT}, 2007.

\bibitem{ketchum:01}
D.~Wang, P.~Jiang, J.~Kukula, Y.~Zhu, T.~Ma, and R.~Damiano.
\newblock {\bf Formal property verification by abstraction refinement with
  formal, simulation and hybrid engines}.
\newblock In {\em DAC}, 2004.

\end{thebibliography}
\end{document}